\documentclass[prl,twocolumn,superscriptaddress]{revtex4-1}

\usepackage{graphicx}
\usepackage{amsmath,amssymb,amsthm,mathrsfs,amsfonts,dsfont,amstext}
\usepackage{textcomp,pbox}
\usepackage[colorlinks=true]{hyperref}
\usepackage{graphicx}
\usepackage{dcolumn,booktabs,url}
\usepackage{bm}
\usepackage{color}
\usepackage{latexsym}
\usepackage{appendix}
\usepackage{threeparttable}
\usepackage{multirow}
\usepackage{makecell}

\usepackage{pifont}
\newcommand{\xmark}{\ding{55}}

\hypersetup{
     colorlinks   = true,
     citecolor    = blue,
     linkcolor    = blue,
     urlcolor     = blue     
}

\begin{document}
\title{Experimental quantum non-Gaussian coincidences of entangled photons}
\newcommand{\USTCmicro}{Hefei National Research Center for Physical Sciences at the Microscale and School of Physical Sciences, University of Science and Technology of China, Hefei 230026, China}
\newcommand{\USTCCAS}{Shanghai Research Center for Quantum Science and CAS Center for Excellence in Quantum Information and Quantum Physics, University of Science and Technology of China, Shanghai 201315, China}
\newcommand{\USTCNational}{Hefei National Laboratory, University of Science and Technology of China, Hefei 230088, China}
\newcommand{\USTCPhy}{School of Physical Sciences, University of Science and Technology of China}
\newcommand{\Palacky}{Department of Optics, Palacký University, 17. listopadu 12, 77146 Olomouc, Czech Republic}
\newcommand{\simit}{Shanghai Key Laboratory of Supercoductor Integrated Circuit Technology, Shanghai Institute of Microsystem and Information Technology, Chinese Academy of Sciences, Shanghai 200050, China}
\author{Run-Ze Liu}
\thanks{These authors contributed equally to this work.}
\affiliation{\USTCmicro}
\affiliation{\USTCCAS}
\author{Yu-Kun Qiao}
\thanks{These authors contributed equally to this work.}
\affiliation{\USTCmicro}
\affiliation{\USTCCAS}
\author{Luk\'{a}\v{s} Lachman}
\thanks{These authors contributed equally to this work.}
\affiliation{\Palacky}
\author{Zhen-Xuan Ge}
\thanks{These authors contributed equally to this work.}
\affiliation{\USTCmicro}
\affiliation{\USTCCAS}
\author{Tung-Hsun Chung}
\affiliation{\USTCNational}
\author{Jun-Yi Zhao}
\affiliation{\USTCmicro}
\affiliation{\USTCCAS}
\author{Hao Li}
\affiliation{\simit}
\author{Lixing You}
\affiliation{\simit}
\author{Radim Filip}
\email{filip@optics.upol.cz}
\affiliation{\Palacky}
\author{Yong-Heng Huo}
\email{yongheng@ustc.edu.cn}
\affiliation{\USTCmicro}
\affiliation{\USTCCAS}
\affiliation{\USTCNational}
\date{\today}
\begin{abstract}
    Quantum non-Gaussianity, a more potent and highly useful form of nonclassicality, excludes all convex mixtures of Gaussian states and Gaussian parametric processes generating them. Here, for the first time, we conclusively test quantum non-Gaussian coincidences of entangled photon pairs with the CHSH-Bell factor $S=2.328\pm0.004$ from a single quantum dot with a depth up to $0.94\pm 0.02$ dB. Such deterministically generated photon pairs fundamentally overcome parametric processes by reducing crucial multiphoton errors. For the quantum non-Gaussian depth of the unheralded (heralded) single-photon state, we achieve the value of $8.08\pm0.05$ dB ($19.06\pm0.29$ dB). Our work experimentally certifies the exclusive quantum non-Gaussianity properties highly relevant for optical sensing, communication and computation. 
\end{abstract}
\maketitle

Non-classical states are of significant importance in quantum information processing. They not only have no counterparts in classical physics but also are necessary for new paradigms for enormous applications, including quantum computation~\cite{wangBosonSampling202019, zhongQuantumComputationalAdvantage2020}, quantum communication~\cite{luMiciusQuantumExperiments2022} and quantum metrology~\cite{ slussarenkoUnconditionalViolationShotnoise2017,nielsen2023deterministic, qinUnconditionalRobustQuantum2023a}. However, not all non-classical states are superior to classical approaches. Gaussian states~\cite{weedbrookGaussianQuantumInformation2012}, whose Wigner function~\cite{wignerQuantumCorrectionThermodynamic1932} in phase-space formalism is a two-dimensional Gaussian function, together with Gaussian operations, which map Gaussian states to Gaussian states, are not sufficient for universal quantum computation~\cite{bartlettEfficientClassicalSimulation2002} and can be efficiently simulated with classical circuits~\cite{mariPositiveWignerFunctions2012}. In contrast, non-Gaussian states and operations, which come from the interaction Hamiltonian beyond quadratic in terms of annihilation and creation operators, are necessary for computation, sensing and long-distance communication~\cite{eisertDistillingGaussianStates2002,duanEntanglementPurificationGaussian2000}.

Thus, the discriminative witnesses for \emph{quantum} non-Gaussian (QNG) states~\cite{walschaersNonGaussianQuantumStates2021}, which rule out Gaussian-state convex hull, are essential. If the investigated quantum state is pure, the necessary and sufficient conditions of quantum non-Gaussianity are the negativity of the Wigner function, as proved by Hudson~\cite{hudsonWhenWignerQuasiprobability1974}, and can be extended to multimode systems~\cite{sotoWhenWignerFunction1983}. However, the correspondence no longer holds for mixed states, e.g., lossy photons (mixed with vacuum states)~\cite{mandilaraExtendingHudsonTheorem2009, walschaersNonGaussianQuantumStates2021}. In Fock-state representation rather than by a phase-space approach~\cite{genoniDetectingQuantumNonGaussianity2013, hughesQuantumNonGaussianityWitnesses2014,parkTestingNonclassicalityNonGaussianity2015}, the first criterion detecting QNG single-photon states beyond -3 dB of loss has been successfully obtained~\cite{filipDetectingQuantumStates2011}. It aims to find the boundary to flexibly exclude the convex hull of Gaussian states using the optimized linear combination of photon-number probabilities. It has been successfully experimentally tested for heralded single-photon sources based on nonlinear optics~\cite{jezekExperimentalTestQuantum2011, strakaQuantumNonGaussianDepth2014a,bauneQuantumNonGaussianityFrequency2014}, quantum dots~\cite{predojevicEfficiencyVsMultiphoton2014,schweickertOndemandGenerationBackgroundfree2018}, atomic ensembles~\cite{davidsonBrightMultiplexedSource2021} and trapped ion~\cite{higginbottomPureSinglePhotons2016}.

However, multimode quantum non-Gaussian photon pairs being crucial resources to advanced quantum communication, sensing and computing, have not been generated yet. Recently, the QNG coincidences criterion~\cite{lachmanQuantumNonGaussianPhoton2021} was proposed to reject the states stemming from Gaussian parametric processes, e.g., spontaneous parametric down-conversion (SPDC). The fluorescence from a solid-state emitter, especially a single quantum dot (QD), is essentially distinct from the SPDC process, which is a Gaussian squeezing operation on the vacuum state. For the single-mode (largely multimode) SPDC process, photons are generated stochastically. The multiphoton probability obeys the Bose-Einstein (Poisson) distribution. In contrast, when an exciton is confined inside a single QD, a pure single-photon Fock state can be, in principle, produced deterministically~\cite{senellartHighperformanceSemiconductorQuantumdot2017a, tommBrightFastSource2021}. When both exciton (X) and biexciton (XX) are confined in a single QD, the cascaded emission can emit a pair of polarization-entangled photons~\cite{huberStrainTunableGaAsQuantum2018a}.

Here, we employ a single QD coupled with an optical microcavity in the weak-coupling regime as an entangled photon source and characterize the deterministically emitted QNG photon pairs. We measure the QNG depth of a single-photon state (see Eq.~\eqref{eqn:QNG_SPS} and Ref.~\cite{strakaQuantumNonGaussianDepth2014a}) up to $8.08\pm0.05$ dB ($19.06\pm0.29$ dB) when one of the entangled photons is unheralded (heralded). More importantly, for the first time, we unambiguously surpass the QNG coincidences criterion (see Eq.~\eqref{eqn:threshold} and Ref.~\cite{lachmanQuantumNonGaussianPhoton2021}) and obtain a depth of $0.94\pm0.02$ dB for the multimode criterion. We note that the values of QNG depths, which characterize the robustness to attenuation, are obtained directly from raw data calculation, without any noise subtraction and loss correction. 

\begin{figure}
    \centering
    \includegraphics[width=1\linewidth]{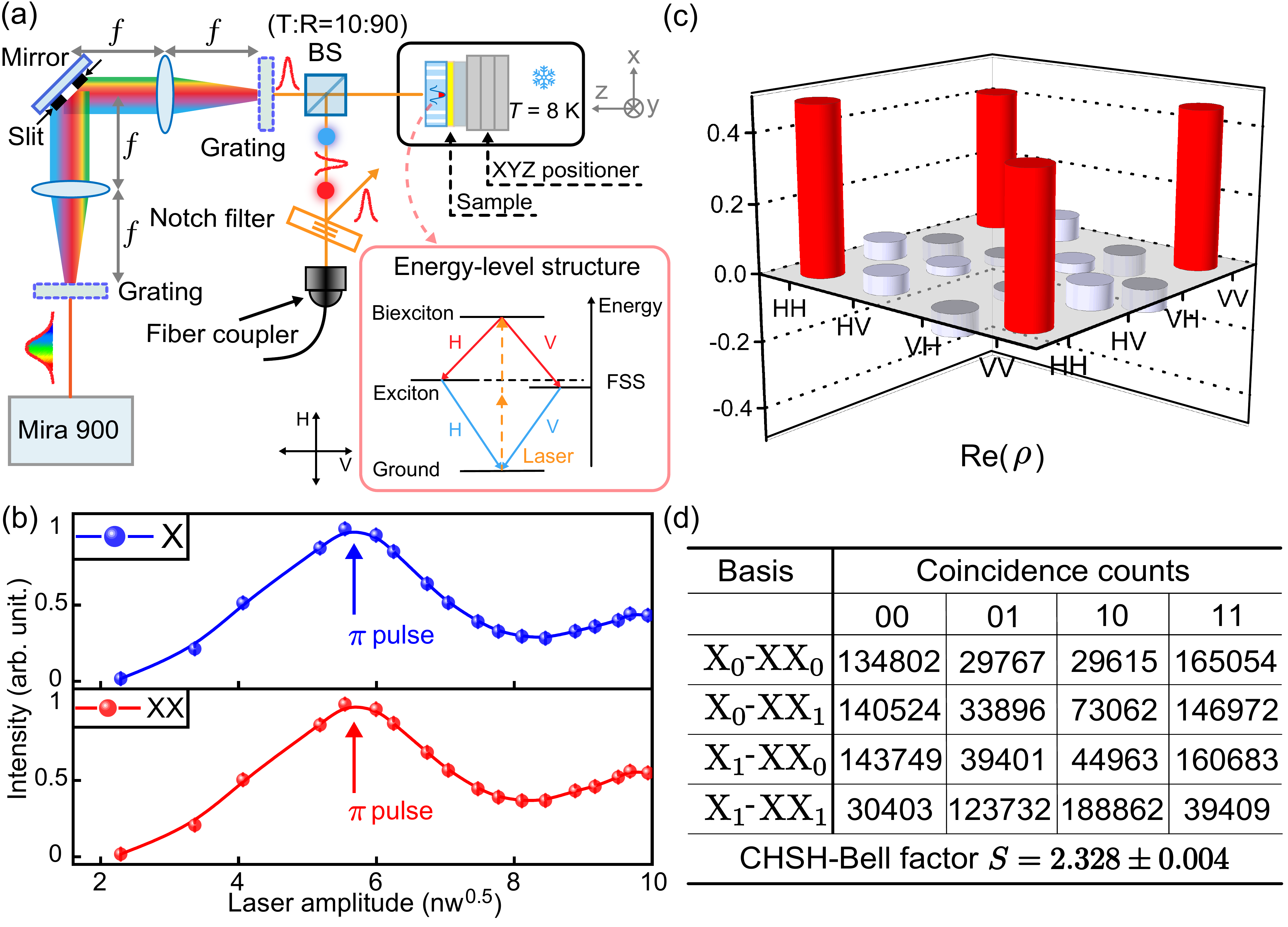}
    \caption{(color online) \textbf{Schematics of QD characterizations.}
    \textbf{(a)} Confocal microscopy setup for QD excitation and fluorescence collection.
    \textbf{(b)} Rabi oscillation confirms the coherent dynamics of two-photon excitation.
    \textbf{(c)} Real part of density matrix for XX-X entangled photon pairs tomography. The calculated fidelity is $90.9\pm0.6$\%.
    \textbf{(d)} CHSH inequality measurement in 10 seconds. The CHSH-Bell factor $S$ is $2.328\pm0.004$.}
    \label{fig:QD}
\end{figure}

In our experiment, we use a GaAs/AlGaAs QD grown by the local droplet etching~\cite{huoUltrasmallExcitonicFine2013b, huoLightholeExcitonQuantum2014a, gurioliDropletEpitaxySemiconductor2019}, integrated with a broadband bull's-eye microcavity~\cite{wangOnDemandSemiconductorSource2019, liuSolidstateSourceStrongly2019a,wangOptimalSinglephotonSources2019} to enhance the extraction efficiency~\cite{yaoDesignHybridCircular2018}. As shown in FIG.~\ref{fig:QD}a, two-photon excitation~\cite{stuflerTwophotonRabiOscillations2006} is used to populate the biexciton state, in which the excitation laser is shaped from a femtosecond laser via a 4$f$-system, and the shaped pulse duration is $5.8$ ps. To confirm the coherent preparation, we gradually increase the laser power and record the intensity of fluorescence. In FIG.~\ref{fig:QD}b, Rabi oscillation can be observed, and the corresponding $\pi$-pulse power is $\sim$ 32 nW. Then, we use quantum state tomography~\cite{jamesMeasurementQubits2001} to reconstruct the density matrix of XX-X entanglement, as shown in FIG.~\ref{fig:QD}c. The calculated fidelity is $90.9\pm0.6$\%, close to the theoretically predicted value of $90.7\%$~\cite{supp}. In entanglement-based quantum cryptography applications, like quantum key distribution (QKD)~\cite{zapateroAdvancesDeviceindependentQuantum2023} and randomness expansion~\cite{liuDeviceindependentRandomnessExpansion2021}, the violation of CHSH inequality~\cite{clauserProposedExperimentTest1969} is the figure of merit. To measure the CHSH-Bell factor $S$, we choose four different basis settings for X and XX photons~\cite{youngBellInequalityViolationTriggered2009}, as shown in FIG.~\ref{fig:QD}d, in which $\text{X}_{0/1}=\sigma_{z/y}$ and $\text{XX}_{0/1}=\frac{(\sigma_z\mp\sigma_y)}{\sqrt{2}}$. $\sigma_{z/y}$ are Pauli matrices and the eigenstates of $\sigma_z$ are H and V polarization in the lab frame. From the results, the violation value can be obtained as $S=2.328\pm0.004$, larger than the classical threshold $2$.

After the characterizations of QD, we examine the first QNG property of fluorescence: unheralded (heralded) single-photon state. The criterion proposed in Ref.~\cite{filipDetectingQuantumStates2011} provides a feasible way to distinguish the single-photon state from a mixture of Gaussian states, even with a positive Wigner function. Furthermore, the depth of the QNG~\cite{strakaQuantumNonGaussianDepth2014a} state was introduced to measure the robustness of the QNG against attenuation. This criterion was previously examined by different physical systems~\cite{jezekExperimentalTestQuantum2011,strakaQuantumNonGaussianDepth2014a,bauneQuantumNonGaussianityFrequency2014,davidsonBrightMultiplexedSource2021,higginbottomPureSinglePhotons2016,predojevicEfficiencyVsMultiphoton2014,schweickertOndemandGenerationBackgroundfree2018}. For solid-state emitters, the brightness and end-to-end efficiency of QD were extremely low to obtain high QNG values with high confidence intervals. For example, in Ref.~\cite{predojevicEfficiencyVsMultiphoton2014}, only 1 coincidence count was recorded within 3 hours. In Ref.~\cite{schweickertOndemandGenerationBackgroundfree2018}, even though the second-order correlation is state-of-the-art, the heralded QNG depth only reaches $5.2\pm1.5$ dB within 10 hours. Until now, the highest unheralded (heralded) QNG depth values of $7.0\pm2.4$ dB (18 dB) were obtained by trapped ion~\cite{higginbottomPureSinglePhotons2016} (heralded SPDC~\cite{strakaQuantumNonGaussianDepth2014a}). Here, we achieve unheralded (heralded) QNG depth up to $8.08\pm0.05$ dB ($19.06\pm0.29$ dB) within 0.5 hours. The discrepancy of unheralded and heralded QNG depth is related to the preparation efficiency and blinking efficiency of QD~\cite{supp}.

\begin{figure}
    \centering
    \includegraphics[width=1\linewidth]{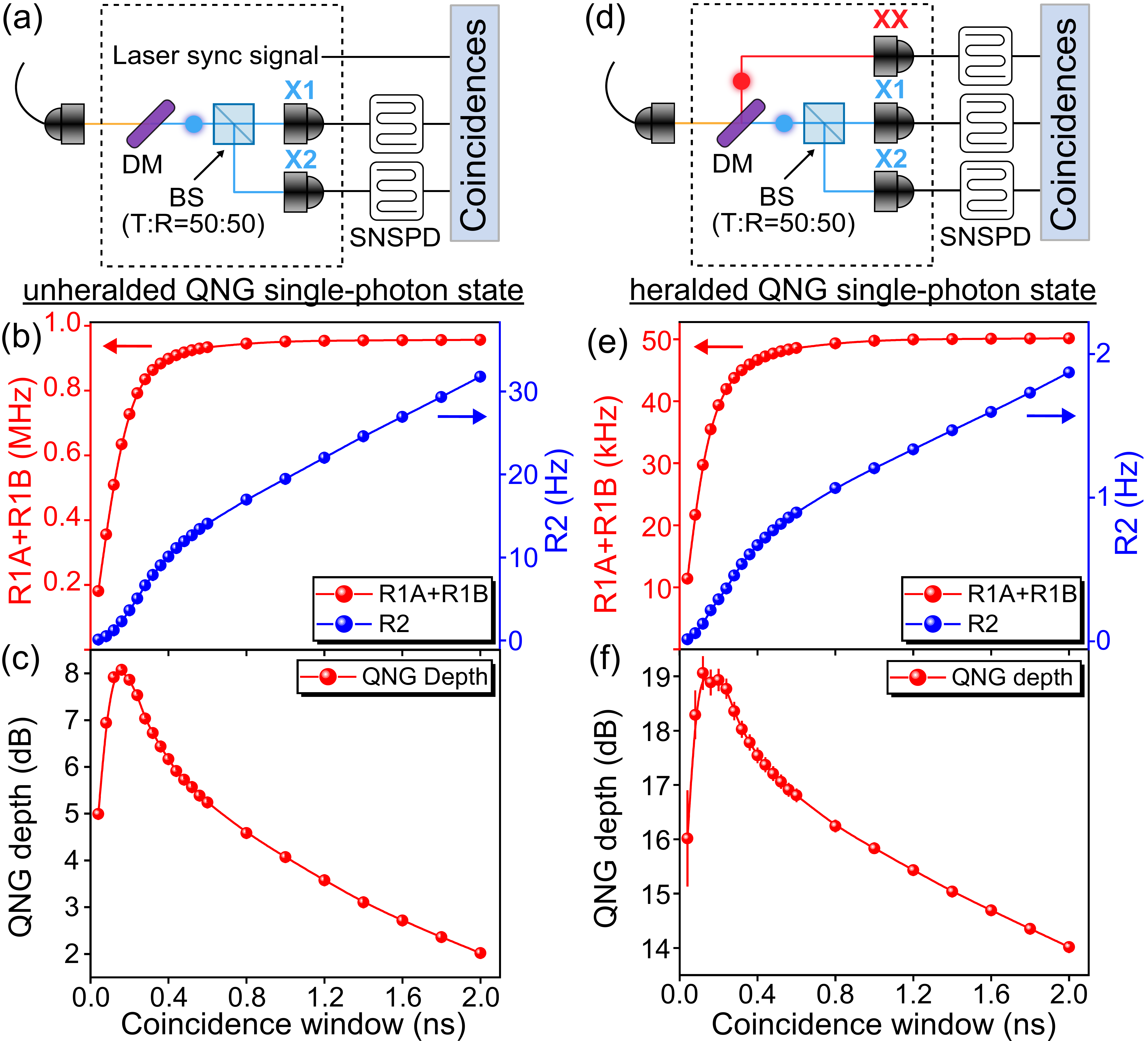}
    \caption{(color online) \textbf{Unheralded and heralded QNG single-photon state.}
    \textbf{(a)} Setup for unheralded QNG single-photon state measurement.
    \textbf{(b)} Unheralded single clicks (R1A and R1B) and unheralded double clicks (R2) coincidence rates at different coincidence windows.
    \textbf{(c)} Unheralded QNG depth at different coincidence windows. The highest QNG depth is $8.08\pm0.05$ dB, when the coincidence window is 0.16 ns.
    \textbf{(d)} Setup for heralded QNG single-photon state measurement.
    \textbf{(e)} Heralded single clicks (R1A and R1B) and heralded double clicks (R2) coincidence rates at different coincidence windows.
    \textbf{(f)} Heralded QNG depth at different coincidence windows. The highest QNG depth is $19.06\pm0.29$ dB, when the coincidence window is 0.12 ns.}
    \label{fig:QNG}
\end{figure}

For the unheralded situation, as shown in FIG.~\ref{fig:QNG}a, the collected photons are split by a dichroic mirror (DM). Then, the X photons are split by a beam splitter (BS) into two spatial modes and detected by commercial NbN superconducting nanowire single-photon detectors (SNSPDs), whose detection efficiency is $\sim 80\%$. The coincidences between the laser sync signal and photons are recorded by a time-to-digital converter (TDC). In the experiment, single clicks are denoted by R1A and R1B for X1-laser and X2-laser coincidences, respectively. R2 represents double clicks, that is, laser-X1-X2 coincidences. The QNG depth of the single-photon state criterion is used to characterize the tolerable attenuation level, at which the QNG property still survives. It can be calculated by the vacuum-state fraction and multiphoton fraction as~\cite{strakaQuantumNonGaussianDepth2014a, jezekExperimentalTestQuantum2011}

\begin{align}
    T_\text{SPS}=-10\log_{10}\left(\frac{3}{2}\frac{P_{2+}}{P_1^3}\right)\ \text{dB} \label{eqn:QNG_SPS}
\end{align}
where $P_{2+}=1-P_0-P_1$ is the multiphoton probability and $P_0, P_1$ are the vacuum and single-photon probabilities, respectively. The single clicks (R1A and R1B) and double clicks (R2) coincidence rates are plotted as functions of the width of coincidence windows, as shown in FIG.~\ref{fig:QNG}b. When the coincidence windows are smaller than 0.8 ns, R1A+R1B and R2 increase cumulatively. After 0.8 ns, the single clicks saturate, but the double clicks grow linearly. This linear trend indicates that the R2 counts are mainly from the flat background noise rather than the imperfection of a single-photon source in large coincidence windows. In the QNG depth calculation, the effect of a slightly unbalanced splitting ratio of BS is also considered~\cite{supp}. As shown in FIG.~\ref{fig:QNG}c, the highest QNG depth is $8.08\pm0.05$ dB when the coincidence window is 0.16 ns. When the coincidence window expands to 0.8 ns, corresponding to 98.3\% of heralded single-photon events, the QNG depth decreases to $4.58\pm0.02$ dB.

As shown in FIG.~\ref{fig:QNG}d, we also measure the heralded situation, in which the XX photons act as heralded signals thanks to cascaded emission. The heralded single clicks and double clicks are X1-XX, X2-XX and X1-X2-XX coincidences, respectively. From FIG.~\ref{fig:QNG}f, the highest QNG depth for the heralded single-photon state is $19.06\pm0.29$ dB when the coincidence window is 0.12 ns. When the coincidence window expands to 0.8 ns, corresponding to 98.2\% of heralded single-photon events, the QNG depth decreases to $16.24\pm0.10$ dB. 

\begin{figure*}
    \centering
    \includegraphics[width=1\textwidth]{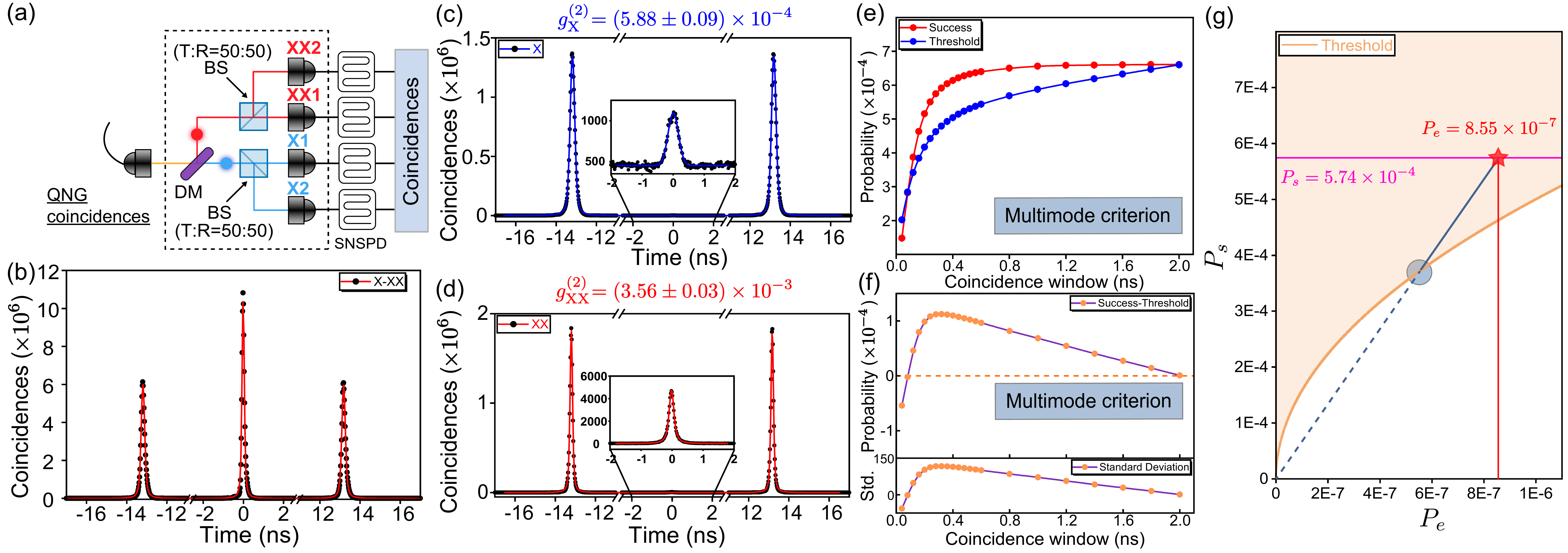}
    \caption{(color online) \textbf{QNG coincidences.}
    \textbf{(a)} Setup for QNG coincidences measurement.
    \textbf{(b)} X-XX correlation. The zero-time peak corresponds to the cascaded emission. 
    \textbf{(c, d)} Second-order correlation of X and XX photons. From fitting, we can deduce the second-order correlation from the ratio of zero-time peak to adjacent peaks as $g^{(2)}_{\text{X}}=(5.88\pm0.09)\times10^{-4}$ and $g^{(2)}_{\text{XX}}=(3.56\pm0.03)\times10^{-3}$.
    \textbf{(e)} Experimental success probability and threshold probability at different coincidence windows.
    \textbf{(f)} Difference between experimental success probability and threshold probability and the corresponding standard deviations at different coincidence windows. When the difference is larger than 0, QNG coincidences property is certified.
    \textbf{(g)} Yellow curve denotes the multimode criterion for QNG coincidences, and the region above this line (yellow shaded) corresponds to the QNG coincidences. In experiment, the best threshold violation is achieved in 0.28 ns (labeled as red star), and the success (error) probability is $5.74\times10^{-4}$ ($8.55\times10^{-7}$). The robustness of QNG coincidences against attenuation is plotted by the blue solid line and the blue shaded circle is the critical point. The QNG coincidences depth $T_\text{coin}$ is $0.94\pm0.02$ dB.} 
    \label{fig:QNG_Coin}
\end{figure*}

QNG coincidences between different modes play an essential role in scalable quantum photonic applications. For example, in entanglement swapping, distributed entangled photon pairs are heralded by the Bell-state analyzer signals, which are the coincidences of interfered photons~\cite{panExperimentalEntanglementSwapping1998, pompiliRealizationMultinodeQuantum2021, hermansQubitTeleportationNonneighbouring2022}. In Ref.~\cite{lachmanQuantumNonGaussianPhoton2021}, a criterion for QNG coincidences is proposed to distinguish the two-photon Fock state $|1\rangle|1\rangle$ in different spatial modes from coincidences provided by a model of a multimode SPDC process. As shown in FIG.~\ref{fig:QNG_Coin}a, both X and XX photons are split by BSs and detected by four SNSPDs. TDC records the coincidences. The criterion is characterized by success probability $P_s$, which quantifies the coincidences from different spatial modes, and error probability $P_e$, which quantifies the coincidences from the same spatial mode~\cite{supp}. For the two-photon Fock state in the different spatial modes, 

\begin{align}
    P_s > \frac{1}{2}\sqrt{P_e}+\frac{3}{8}P_e+\frac{1}{16}P_e^{3/2}
    \label{eqn:threshold}
\end{align}
holds~\cite{lachmanQuantumNonGaussianPhoton2021}. As shown in FIG.~\ref{fig:QNG_Coin}b, the X-XX correlation is plotted, and the zero-time peak corresponds to the cascaded emission. We note that the preparation efficiency can be deduced from the ratio of coincidence counts in other peaks to coincidence counts in the zero-time peak~\cite{wangOnDemandSemiconductorSource2019}. For the investigated QD, the preparation efficiency is $84.7\pm0.6 \%$ under two-photon excitation. The deviation from unity can be attributed to the phonon-induced dephasing process~\cite{ramsayDampingExcitonRabi2010, javadiCavityenhancedExcitationQuantum2023}. The error probability is from the imperfect single-photon purity of X (XX) photons. As shown in FIG.~\ref{fig:QNG_Coin}c and~\ref{fig:QNG_Coin}d, the second-order correlations are $(5.88\pm0.09)\times10^{-4}$ and $(3.56\pm0.03)\times10^{-3}$ for X and XX photons, respectively. The remaining imperfections come from the residual laser and very weak emission from neighbouring states~\cite{supp}.

We can calculate the success probability and the error probability with the coincidence counts and laser repetition rate (75.84 MHz). As shown in FIG.~\ref{fig:QNG_Coin}e, the experimental success and threshold probabilities (calculated with error probability using Eq.~\eqref{eqn:threshold}) are plotted as functions of different coincidence windows. When the coincidence windows enlarge, the success probability increases and finally saturates. The threshold probability increases gradually because more background noise is included. To make it more intelligible, as shown in FIG.~\ref{fig:QNG_Coin}f, the difference between the experimental success and threshold probability is plotted as a function of coincidence windows. When the difference value is positive, the QNG coincidences criterion is surpassed. In the experiment, the best threshold violation is achieved in the 0.28 ns coincidence window. The corresponding success and error probabilities are $5.74\times10^{-4}$ and $8.55\times10^{-7}$ respectively, which exceed the QNG coincidences criterion by more than 116 standard deviations. Similar to the QNG depth for a single-photon state, we also define the QNG coincidences depth as the minimum transmissivity $T$ of a lossy channel that preserves the criterion Eq.~\eqref{eqn:threshold}. Since the lossy channel reduces the probabilities following $P_{s/e}(T)=P_{s/e}T^2$, we can quantify the QNG coincidences depth as

\begin{align}
    T_{\text{coin}}=-10\log_{10}\left(\frac{\sqrt{P_e}}{2P_s}\right)\ \text{dB}
\end{align}
when $P_e$ is negligible. As shown in FIG.~\ref{fig:QNG_Coin}g, because the ratio $P_s(T)/P_e(T)$ is constant, the robustness of the QNG coincidences against attenuation is plotted by the blue solid line, and the QNG coincidences depth is $0.94\pm0.02$ dB. We also note that the criterion of Eq.~\eqref{eqn:threshold} is used for ruling out the most common Gaussian SPDC process. Ref.~\cite{lachmanQuantumNonGaussianPhoton2021} also derives another criterion in which every photon in a single temporal mode is essential for distinguishing from a convex mixture of \emph{all} two-mode Gaussian states. For QD entangled photon source, due to temporal correlation, the first-order coherence and indistinguishability of photons are inhibited by the ratio of X and XX lifetimes~\cite{simonCreatingSingleTimeBinEntangled2005b, schollCruxUsingCascaded2020a,supp}. For better indistinguishability of photons, quantum interferences can be used to eliminate the temporal correlation restriction and obtain coherent polarization-entangled photon pairs~\cite{liu2023eliminating}.

\begin{table}[ht!]
    \centering
    \caption{Comparison between different QNG experiments}
    \label{tab:comparion}
    \begin{threeparttable}
    \begin{tabular}{c|c|c|c|c}
    \hline
    Reference & \makecell[c]{Physical \\ system} & \makecell[c]{unheralded \\ $T_\text{SPS}$ (dB)} & \makecell[c]{heralded \\ $T_\text{SPS}$ (dB)} & $T_\text{coin}$ (dB) \\ \hline
    Ref.~\cite{strakaQuantumNonGaussianDepth2014a} & SPDC & \xmark & 18 & \xmark \\ \hline
    Ref.~\cite{higginbottomPureSinglePhotons2016} & ion & $7.0\pm2.4$\tnote{a} & \xmark & \xmark \\ \hline
    Ref.~\cite{predojevicEfficiencyVsMultiphoton2014} & QD & \xmark & $5.6\pm4.3$ & \xmark \\ \hline
    Ref.~\cite{schweickertOndemandGenerationBackgroundfree2018} & QD & \xmark & $5.2\pm1.5$ & \xmark \\ \hline
    Our work & QD & $8.08\pm0.05$ & $19.06\pm0.29$ & $0.94\pm0.02$ \\
    \hline
    \end{tabular}
    \begin{tablenotes}
        \item[a] Calculated from digitalized FIG. 4 in paper~\cite{higginbottomPureSinglePhotons2016}.
    \end{tablenotes}
    \end{threeparttable}
\end{table}

To highlight the progress entailed by our results, we compare the results with other QNG experiments, as shown in Table~\ref{tab:comparion}. In summary, using a single QD entangled photon source coupled with an optical microcavity, we achieve the value of $8.08\pm0.05$ dB ($19.06\pm0.29$ dB) for the QNG depth of the unheralded (heralded) single-photon state. Though the results outperform measured values in the literature, we also note the state-of-the-art single-photon source based on QD~\cite{ding2023highefficiency,tommBrightFastSource2021} or single atom~\cite{shi2022high,daiss2019single} might obtain higher QNG depth. Besides, for the first time, we unambiguously surpass the QNG coincidences criterion, which distinguishes the photon pair generation from the two-mode squeezing state, and we measure the QNG coincidences depth up to $0.94\pm0.02$ dB. All QNG depths are calculated directly from raw data, without any noise subtraction. With precise QD-cavity coupling to increase the overall efficiency~\cite{ding2023highefficiency,tommBrightFastSource2021, najerGatedQuantumDot2019a}, a substantial improvement in the QNG depth is foreseeable~\cite{supp}. A stringent QNG coincidences criterion, highlighting the exclusive properties of quantum states embedded in a subspace with largely suppressed multiphoton errors, can be new standards for developing quantum light sources. For example, a boson sampler necessitates a train of efficient and pure single photons~\cite{wangBosonSampling202019}. Besides, the QNG criteria can assess the scalability of quantum communication applications, including the achievable distribution scale of photons~\cite{weberTwophotonInterferenceTelecom2019a,10.1117/1.AP.4.6.066003,schimpfQuantumCryptographyHighly2021}, indicating security of QKD~\cite{lasota2017sufficiency}, and the advantages of realistic QD photon sources against Poisson-distributed sources~\cite{bozzioEnhancingQuantumCryptography2022}.

\begin{acknowledgements}
R.-Z. L., Y.-K. Q., Z.-X. G., T.-H. C., J.-Y. Z., and Y.-H. H. gratefully acknowledge financial support from Innovation Program for Quantum Science and Technology (Grant No. 2021ZD0300204), Shanghai Municipal Science and Technology Major Project (Grant No.2019SHZDZX01), Anhui Initiative in Quantum Information Technologies, National Natural Science Foundation of China (Grant No. 11774326) and National Key R\&D Program of China (Grant No. 2017YFA0304301). R. F. and L. L. acknowledge the support of project No. 21-13265X of the Czech Science Foundation and the European Union’s 2020 research and innovation programme (CSA - Coordination and support action, H2020-WIDESPREAD-2020-5) under grant agreement No. 951737 (NONGAUSS).
\end{acknowledgements}

%

\end{document}